# McMini: A Programmable DPOR-Based Model Checker for Multithreaded Programs

Maxwell Pirtle[a], Luka Jovanovic[a], and Gene Cooperman[a]

a   Khoury College of Computer Sciences, Northeastern University, United States, Boston, USA

**Abstract**
**Context**   Model checking has become a key tool for gaining confidence in correctness of multi-threaded programs. Unit tests and functional tests do not suffice because of race conditions that are not discovered by those tests. This problem is addressed by model checking tools. A simple model checker is useful for detecting race conditions prior to production.
**Inquiry**   Current model checkers hardwire the behavior of common thread operations, and do not recognize application-dependent thread paradigms or functions using simpler primitive operations. This introduces additional operations, causing current model checkers to be excessively slow. In addition, there is no mechanism to model the semantics of the actual thread wakeup policies implemented in the underlying thread library or operating system. Eliminating these constraints can make model checkers faster.
**Approach**   McMini is an *extensible* model checker based on DPOR (Dynamic Partial Order Reduction). A mechanism was invented to declare to McMini new, primitive thread operations, typically in 100 lines or less of C code. The mechanism was extended to also allow a user of McMini to declare alternative thread wakeup policies, including spurious wakeups from condition variables.
**Knowledge**   In McMini, the user defines new thread operations. The user optimizes these operations by declaring to the DPOR algorithm information that reduces the number of thread schedules to be searched. One declares: (i) under what conditions an operation is enabled; (ii) which thread operations are independent of each other; and (iii) when two operations can be considered as co-enabled. An optional wakeup policy is implemented by defining when a wait operation (on a semaphore, condition variable, etc.) is enabled. A new enqueue thread operation is described, allowing a user to declare alternative wakeup policies.
**Grounding**   McMini was first confirmed to operate correctly and efficiently as a traditional, but extensible model checker for mutex, semaphore, condition variable, and reader-writer lock. McMini's extensibility was then tested on novel primitive operations, representing other useful paradigms for multithreaded operations. An example is readers-and-two-writers. The speed of model checking was found to be five times faster and more, as compared to traditional implementations on top of condition variables. Alternative wakeup policies (e.g., FIFO, LIFO, arbitrary, etc.) were then tested using an enqueue operation. Finally, spurious wakeups were tested with a program that exposes a bug *only* in the presence of a spurious wakeup.
**Importance**   Many applications employ functions for multithreaded paradigms that go beyond the traditional mutex, semaphore, and condition variables. They are defined on top of basic operations. The ability to directly define new primitives for these paradigms makes model checkers run faster by searching fewer thread schedules. The ability to model particular thread wakeup policies, including spurious wakeup for condition variables, is also important. Note that POSIX leaves undefined the wakeup policies of `pthread_mutex_lock`, `sem_wait`, and `pthread_cond_wait`. The POSIX thread implementation then chooses a particular policy (e.g., FIFO, arbitrary), which can be directly modeled by McMini.

**ACM CCS 2012**
- **Software and its engineering** → **Formal methods**;

## The Art, Science, and Engineering of Programming

### McMini: A Programmable DPOR-Based Model Checker for Multithreaded Programs

## 1 Introduction

McMini is a free and open-source model checker of about 5,000 lines of code (not including its test suite). It is based on DPOR (Dynamic Partial Order Reduction) [17] (see Section 2 for background). Its small size and object-oriented design allows others to modify it. Yet its implementation includes common performance optimizations suggested in the original DPOR paper, such as sleep sets and clock vectors.

The novelty arises from the extensibility or programmability of McMini, which allows others to: (a) define new multithreaded constructs; (b) to define alternative wakeup policies; and (c) to enable spurious wakeups when using condition variables. Defining a new thread operation (and a wakeup policy if it involves a "wait" operation) typically requires 100 lines of C code or less. McMini diagnoses deadlock, assertion failures and segfaults (e.g., due to memory corruption). It also includes heuristics to check for data races (see Section 5.8).

Model checking has become a key tool for gaining confidence in correctness of multi-threaded programs. For single-threaded programs, unit tests and functional tests are widely used. But for multi-threaded programs these techniques often miss important race conditions. There are many sophisticated model-checking tools for multi-threaded programs available [11, 29, 34]. However, those existing tools either require significant experience to use effectively [15, 18, 26, 27, 31, 33]; or they rely on "hardwired" support for the most common multi-threaded synchronization constructs [7, 44]; or else both, as in the examples of Gambit [13], Godefroid [21], CHESS [23, 33], and the S4U interface [37] of SimGrid [9].

Further, existing model checkers often lack direct support for the less common multi-threaded synchronization constructs, such as futexes, recursive mutexes, robust mutexes, barriers, variants of read-write locks with alternative policies (e.g., reader-preferred or writer-preferred), and condition variables *in the presence of spurious wakeups*.

In current practice, a multithreaded application may build a specialized operation, such as reader-preferred reader-writer locks, on top of simpler primitives such as condition variables. Existing model checkers then analyze the special operation in terms of the supported operations used to define the special operation. However, such implementations are inefficient, since they require multiple calls to lower-level primitives for each call to the specialized operation. As a result, additional thread schedules are needlessly searched, although they do not alter the original semantics.

Similarly, there are instances when a user of McMini would like to declare to the model checker a restricted use case for some well-known thread operation for efficiency reasons. For example, when launching a child thread, it is common to employ a semaphore to synchronize parent and child, in such a way that the semaphore's internal value can only take on values 0 and 1, and other values are an error. (For example, see [6, Section 32.2]: "Order-Violation Bugs".) Another example of a restricted use case is readers-and-two-writers (see Section A.3).

In McMini, a user implements a restricted use case by "editing" McMini's definition of the semaphore or reader-writer lock, and adapts it to define the variant semaphore or reader-writer lock. This typically greatly reduces the search space of alternative thread schedules. This captures better the application's intended semantics, and results in a more efficient DPOR-based model checker.

Beyond the need to define additional multi-threaded constructs, one may also wish to define alternative wakeup policies. Such operations as pthread_mutex, sem_wait, pthread_cond_wait place a thread in a sleep state. Current practice is to allow an *arbitrary* thread to "wake up" when a signal is sent, a mutex is unlocked, or a semaphore is posted to. This is also the *default* policy of McMini.

But the arbitrary wakeup policy often does not reflect the underlying semantics of the thread library, which might use a FIFO queue or other policy. McMini allows the user to implement alternative wakeup policies, such as FIFO.

McMini allows a user to define particular wakeup policies, by adding a McMini "enqueue" operation. The "enqueue" is executed before the thread's McMini "wait" operation, such as sem_wait, pthread_cond_wait, and pthread_mutex_lock. Taking semaphores as an example, a call to sem_post can result in either of two distinct behaviors: (i) Maybe the post arrived before executing the call to "wait", and the wakeup policy will choose a different thread to wake up. (ii) Or maybe the post arrived after the call to "wait" (and before the "wait" could complete), and the waiting thread is eligible to be woken up. See Section 4.3 for a detailed discussion on the enqueue operation, and Section 5.6 for experiments showing the effects of different policies.

Versions of McMini were used in the classroom by the author Cooperman to teach multi-threaded programming. This allowed students to go beyond the standard "bank account", and "consumer-producer" programs, and reason about the logic and potential races in a large range of multi-threaded programs. This disabused the students of the notion that race conditions in multi-threaded programs could be uncovered through test suites, unit tests, and verifying code "by eye".

"Buggy" programs with race conditions were presented to the students, and they were asked to analyze the bugs, optionally with the assistance of McMini. They were then given a small reader-writer problem with an unusual specification, where they had the opportunity to use McMini to learn from mistakes by gaining immediate feedback, which identified potential race conditions in their code.

An early version of McMini was used previously in an undergraduate course. However, it used a naive model-checking algorithm that lacked DPOR support and hence was quite slow. Nevertheless, the student reception was sufficient to inspire the faster version of McMini presented here, including optimizations for DPOR, sleep sets, and clock vectors. The faster version was then used for the M.S. course and Ph.D. course. In the future, the programmability of McMini will allow students to explore even more subtle issues, such as spurious wakeup for condition variables.

The McMini model checker demonstrates several points of novelty:

1. The user can define new multi-threaded constructs or thread operations.
2. For thread "wait" operations, the user can define alternative wakeup policies closer to the policies of the underlying thread library.
3. A traditional debugger, such as GDB, can be used to examine a particular thread schedule — perhaps one leading to deadlock.

4. Spurious wakeups (e.g., in condition variables) are supported, in which the use configures the maximum number of spurious wakeups for a given condition variable.

The remainder of this work includes some background on DPOR (Section 2), the McMini software architectures (Section 3), the API for creating new thread primitives (Section 4), an experimental evaluation (Section 5), related work (Section 6), and a conclusion along with future work (Section 7).

## 2 Background: DPOR Viewed as a Meta-programming Framework

The original DPOR (dynamic partial order reduction) technique for model checking was developed in a landmark paper by Flanagan and Godefroid [17]. We define here some terminology for the basic concepts. Details of DPOR, along with more formal definitions of the terminology can be found in that same landmark paper [17].

Intuitively, DPOR, like any model checking algorithm, explores all possible interleavings of executions of the threads. The operations executed by a thread can be divided into *invisible operations* (operations that have no effect on the other threads), and *visible operations* (thread operations: operations that can eventually affect another thread). We are worried only about interleavings among the visible operations, since interleavings among the invisible operations will have no effect on the final computation. For this reason, a *transition* (one interleaved step in a program) consists of execution by a particular thread that consists of a single *visible operation* followed by arbitrarily many *invisible operations*, until reaching and stopping before the next visible operation. Further, a *visible operation* must be an *atomic operation*, that cannot be split by further interleavings. Examples of visible operations include operations on mutexes, semaphores and condition variables — but also read and write operations on variables that are shared among the threads.

Note that each multithreaded execution is equivalent to some interleaving or sequential version of that program according to a *thread schedule*, in which the program is executed sequentially in a series of steps, and each step consists of one thread executing one transition.

When we "freeze" the process at some step of a thread schedule, then each thread has the possibility of executing next in the thread schedule. A thread is *enabled* if it can complete the visible operation associated with its next transition. For example, if a thread's next transition would begin with executing the visible operation pthread_mutex_lock(&mutex), then the thread is enabled if and only if no other thread holds the lock associated with the variable mutex. A thread is *not enabled* if it cannot complete the visible operation associated with its next transition.

Continuing to consider this "frozen" process, we consider two threads that are both enabled and may execute the next step. We say that the next transitions for each of two threads are *independent* if we can schedule the threads in either order to reach the same state. For example, if two threads each intend to execute transitions beginning with a visible operation sem_post(&semaphore) on the same semaphore, then the two transition are independent. Those steps can execute in either order to reach the same state of the program. The transitions for two threads are *dependent* if they are not independent.

*Pruning* of search branches is the key intuition behind the elaborate depth-first search strategy offered by the DPOR algorithm. A simple example of *pruning* is given next. Assume the next transitions for each of thread A and thread B are independent. Then we could execute either the interleaving of A followed by B or of B followed by A, and the result is the same. In this case, it suffices to ignore all interleavings that execute a transition of B followed by A, as long as we search all transitions beginning with A and followed by B.

Finally, two visible operations are *co-enabled* if there exists some process state (a "frozen" process according to some thread schedule) such that the visible operation of the next transition for each of two threads are both enabled. More loosely, two visible operations are co-enabled if there exists some process state where those two visible operations are part of transitions for two threads, and both transitions are enabled in this process state. For example, pthread_mutex_lock(&mutex) and pthread_mutex_lock(&mutex) (the same visible operation on the same mutex) are co-enabled, but pthread_mutex_lock(&mutex) and pthread_mutex_unlock(&mutex) are not co-enabled. Either mutex is locked or it is not locked: there is no single case where we could either lock or unlock the same mutex.

The second attribute (deciding when two visible operations are *independent*) is the essence of the DPOR-based optimization. In its simplest form, a search can be pruned upon determining that the next visible operations by thread A and thread B are *independent*, and hence they do not interact. One can schedule thread A followed by thread B or vice versa, and the process state will be the same. So, only one schedule needs to be searched. The DPOR algorithm goes further, and examines entire subsequences of the visible operations in a thread schedule. In doing so, it maintains "backtrack points", and can prune branches from much earlier backtrack points.

The first and third attributes are not as closely associated with DPOR optimizations. The first attribute (when a visible operation is enabled) is common to any model-checking software for threads — whether based on DPOR or not. This simply determines which threads are eligible to be scheduled next in a depth-first search of all possible thread schedules. The third attribute (when visible operations are co-enabled) is a minor optimization in DPOR that in some situations removes the need to directly test if a thread is enabled.

## 3 Overview of McMini

McMini employs the DPOR technique. McMini generates thousands of program traces or more in a dynamic analysis, by directly executing the target program for each trace. McMini assumes the user-supplied model for a new visible operation to be correct, and does not directly check the model's correctness. McMini also assumes that external interactions such as file I/O and user input are deterministic for the case being tested. The user executes the program for a fixed input and files, chosen by the user.

## 3.1 The McMini Software Architecture

McMini can be invoked as simply as: mcmini ./a.out … . McMini is implemented by calling exec() on ./a.out, and preloading the libmcmini.so library during exec(), using the LD_PRELOAD environment variable.

The core novelty of McMini, is the ability to "program" a new multi-threaded construct. Each new multi-threaded operation is a "visible operation", according to the above terminology. In order for McMini to integrate the visible operation into the DPOR algorithm, the user must write short routines to define:

1. when a transition for this visible operation is *enabled*;
2. when transitions for this visible operation and another are *independent*; and
3. which other visible operations are *co-enabled* with this visible operation.

The "scheduler" portion of the library contains logic for controlling the execution of threads in the application, as well as implementations for each transition that McMini supports. The libmcmini.so library also contains definitions (or wrapper functions) for each of the (dynamically-linked) functions in the application that should be treated as visible operations (viz. for each transition defined in the scheduler). McMini makes the assumption that any visible operation within the application ultimately results in a call to a function residing in a dynamic library (e.g., libpthread.so), so that wrapper functions can be implemented. This allows McMini to perform model-checking without modification of the target executable or its libraries.

The scheduler portion of libmcmini.so repeatedly forks itself for each successive thread schedule. A schedule runs the first thread whose next transition is enabled, updating the DPOR-based backtracking sets as it proceeds forward. When there are no threads whose next transition is enabled, McMini finds the most recent backtracking point and repeats the process. Figure 1 illustrates the basic software architecture.

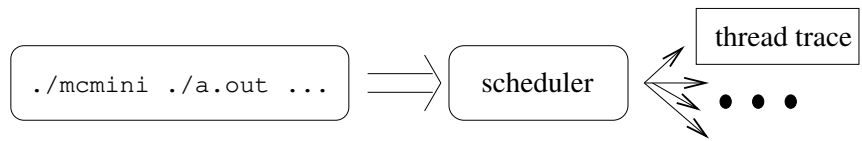

■ **Figure 1** For a multithreaded application a.out, the user types ./mcmini ./a.out …. McMini then exec's into the program ./a.out …, but also preloads a libmcmini.so library. The library takes control first, through a constructor function, and then forks a child process for each thread trace to be tested by DPOR. The thread trace executes a.out, but stops in a wrapper function around each thread operation. Within the wrapper, the thread trace communicates back to the scheduler. The scheduler uses the DPOR algorithm to decide which thread to schedule next in the current thread trace. Any file I/O is assumed to be deterministic.

McMini supports several thread primitives out-of-the-box, along with the common operations acting on them, including pthread_t, pthread_mutex_t, pthread_rwlock_t, sem_t, pthread_cond_t,and pthread_barrier_t.

### 3.2 Scheduling Behavior

McMini executes a depth-first search with a maximum depth in the number of transitions in a trace. For each trace, McMini forks a child process and re-runs the threads in the program up until the chosen backtracking point and then extends the trace, extending the new trace first with the transition run by the thread popped from the backtracking set. The forked child processes serve only to report to the scheduler process the new transitions executed by each thread in the target program.

By default, McMini continues to run thread traces according to the DPOR algorithm until it has exhausted all DPOR backtracking points. When it reaches a deadlock in the target program (when no thread is enabled at the end of a certain thread schedule or trace), it reports the trace and continues exploring other traces from a different backtrack point. McMini can be configured to stop at the first deadlock and, additionally, restrict the search to a maximum depth per thread ($N$ transitions), which is consistent with strategies of other context-bounded model-checkers [32, 35].

When McMini is configured to restrict the search to a maximum depth per thread, McMini artificially marks threads which have run for more than the thread execution limit as disabled. McMini tracks which threads have exhausted their budget to distinguish between true deadlock and having executed all threads to their limit.

It is important that *each* thread be allowed to run for $N$ transitions, for $N$ sufficiently large, in order to ensure correctness. Restricting each thread to $N$ transitions effectively creates a modified version of the target program, whose threads run up to $N$ transitions in the target and then exit. DPOR guarantees that all deadlocks that can be reached will be reached, within this budget of $N$ transitions per thread. As expected, McMini does not report assertion failures that occur more than $N$ *total* steps into a trace.

### 3.3 Wrapper Functions: Interposing on Thread Primitives

For each visible operation McMini intercepts, McMini's own (re)definition contains custom logic that coordinates with the "scheduler" portion of libmcmini.so. In most cases, this simply means executing the following sequence:

1. Writing, to a predefined portion of shared memory, both the type of transition that was executed, along with a payload specific to that transition (e.g. an address identifying a visible object in the application). This lets the scheduler know which transition the executing thread is *about* to execute.
2. Notifying the scheduler that the calling thread has finished executing its last transition and then blocking until scheduled for execution again

   Each thread in the target process is assigned a unique integer id and a pair of binary "scheduling" semaphores in shared memory and are used by McMini internally: one for the thread in the forked process to sleep on and one for McMini's scheduler to wait on. Each thread initially calls `sem_wait` on its own scheduling semaphore; when McMini schedules a thread to run, a `sem_post` is issued to that thread's semaphore. After reaching the next visible operation, the scheduled thread wakes

McMini with a sem_post on McMini's semaphore and again calls sem_wait on its own scheduling semaphore, sleeping until McMini decides to run that thread again.

When the scheduler is notified of the completion of a transition executed by the last scheduled thread, McMini inspects the type information written in step (1) and invokes a handler mapped to that type that converts the low-level information written by the wrapper into an MCTransition object. The handler is also responsible for creating any new MCVisibleObjects to represent any new visible objects McMini should know about that the application manipulates.

3. Executing the true underlying wrapped function and exiting the wrapper function

   McMini executes the actual multithreaded constructs in the glibc or other thread library whenever possible, except to support different thread orderings for sem_t and pthread_cond_t

### 3.4 GDB Interface to McMini

When a user of McMini has found a trace with a deadlock, or a race leading to an assertion failure or crash, the next step is to examine the thread schedule that leads to the bug. In order to assist the user, McMini has extended the standard GDB debugger with additional commands.

For example, after McMini has reported a segfault for a.out at trace 123, one executes: mcmini-gdb ./a.out . In a typical scenario, one might see:

■ **Listing 1** Example scenario for mcmini-gdb

```
1  (gdb) mcmini gotoTrace 123
2  *** trace: 123; transition: 0; thread: 3.1 (thread 1 of inferior 3)
3  (gdb) mcmini forward 15
4  *** trace: 123; transition: 15; thread: 3.2 (thread 2 of inferior 3)
5  (gdb) mcmini back
6  *** trace: 123; transition: 14; thread: 4.2 (thread 2 of inferior 4)
7  (gdb) # The usual GDB commands are still available: next, step, print x, etc.
```

## 4  The McMini API for Creating New Thread Primitives

The McMini API is illustrated in three subsections. First, Section 4.1 provides an overview of the object-oriented methods needed to define a new thread operation (a new type of transition). Section 4.2 presents the lines of code necessary to implement the thread primitives associated with each of those thread abstractions.

Finally, Section 4.3 describes how to add an *enqueue* operation for semaphores. As described there, any operations that can cause a thread to wait must be split into two steps for proper modeling. The two steps correspond to "entering" the function (changing the state of that thread of that thread to "waiting"), and the second step corresponds to "exiting" the function (changing the state of that thread to "running").

Note that McMini uses its API both to implement its own built-in primitives, and to create new thread primitives. McMini's built-in primitives include support for mutex, semaphore, condition variable, barrier, and for the basic thread abstraction (pthread_create(), pthread_join()).

### 4.1 Creating a New Thread Primitive

All McMini objects are defined in the same way. One begins by extending the McMini class MCVisibleObject. Each object maintains information to capture the correct semantics of the primitive it describes. Objects also provide methods to query the object's state for access inside the transitions that operate on them. McMini creates objects when first encountered and persist for the duration of model checking. Example objects include MCMutex, MCSemaphore, and MCConditionVariable.

Recall from Section 2 that the goal of DPOR is to *prune* additional branches in a model checker based on depth-first search. DPOR-style search requires that for each visible operation, three attributes must be defined: enabled, independent, and coenabled. McMini implements each new transition (i.e., a visible (multi-threaded) operation) as an extension of an MCTransition. An MCTransition has three default methods, which must be extended.

1. bool enabledInState(const MCState *) — must be overridden
2. bool dependentWith(const MCTransition *) — default is true
3. bool coenabledWith(const MCTransition *) — default is true

A transition maintains and operates on references to a MCVisibleObject that is managed by the MCState. A transition is paired with one (or more) wrapper functions, which transparently intercepts calls made in the application to coordinate with McMini.

Example transitions include MCMutexLock, MCMutexUnlock, MCSemWait, MCSemPost, MCCondWait, and MCCondSignal. In one example, the constructor for a MCMutexLock takes both a mutex and the thread acquiring the mutex as arguments. And as expected, the MCMutexLock defines the three methods of MCTranstion referred to above. One example is:

bool MCMutexLock::enabledInState(const MCState *state)

McMini handles "common" cases for among transitions to reduce the amount of boilerplate code needed to write the special functions in the MCTransition struct. In particular, the special functions can be implemented under the assumption that:

1. the other transition is run by a different thread; and
2. the other transition does not create or join on the thread running this transition.

Further, a transition only needs to specify its own dependencies. McMini will check for a dependency between any two transitions by asking each transition whether it is dependent with the other. If *either* transition specifies that it's dependent with the other, the transitions are considered to be dependent with one another *even if* only one of the two transitions claimed dependence with the other. Thus, one can write any dependency conditions assuming only the existence of any transition subclasses present when implementing the function, and it will be up to future subclasses to properly take into consideration any dependencies with the transitions you define. The same is true when defining whether a transition is co-enabled with another.

The result is a flexible system in which dependencies and co-enabled relations, once written, need not be changed as more transition types are added to McMini.

McMini also defines the `MCState` struct (class). It describes a snapshot in time of a program as DPOR probes it. It contains both the state needed to implement the DPOR algorithm as well as what it believes to be the current state of the program (what each thread is about to execute, what the states of the visible objects are, etc). It also includes methods for adding a new transition, for applying/reverting the effects of a transition when needed for backtracking, for creating a new thread, and various other public auxiliary methods.

As the DPOR algorithm is executed and transitions are taken in the application undergoing testing, McMini notifies its `MCState` about those changes. McMini can then query the `MCState` object to determine how to proceed with the DPOR search.

As an example, the wrapper function for `pthread_mutex_lock` is shown in Listing 2. A new wrapper function is written for each new transition. When the target application executes `pthread_mutex_lock`, McMini arranges for it to call `mc_pthread_mutex_lock`. An `MCMutexShadow` is created to capture the identity of the locked mutex. Then, `thread_post_visible_operation_hit` carries out the relevant bookkeeping of this wrapper call. Next, the `thread_await_scheduler` routine yields control to the McMini scheduler process. Eventually, the thread executing this process will be scheduled again. When this happens, the wrapper executes `__real_pthread_mutex_lock`, which is aliased to the native `pthread_mutex_lock` function (typically defined in libpthread.so). This ensures that McMini is faithful to the native implementation of locks. Finally, the wrapper function returns; and since the current thread retains control, execution continues until `thread_await_scheduler` is again called from some wrapper function.

■ **Listing 2** A wrapper function for pthread_mutex_lock()

```
1  mc_pthread_mutex_lock(pthread_mutex_t *mutex)
2  { auto mutexShadow = MCMutexShadow(mutex);
3    thread_post_visible_operation_hit<MCMutexShadow>(typeid(MCMutexLock),
4      &mutexShadow);
5    thread_await_scheduler();
6    return __native_pthread_mutex_lock(mutex);
7  }
```

## 4.2 The Lines of Code to Implement a New Thread Primitive

Adding new multi-threaded primitives and operations is straightforward. It generally requires less than 100 lines of code to add a thread operation, as seen in Table 1.

A running example using these principles to implement `pthread_rwlock_rdlock()` is provided in the Appendix A.

■ **Table 1** Lines of code (LOC) needed to support each thread operation. pthread_t includes pthread_create and pthread_join.

| Primitive | Visible Operations | LOC | LOC / Operation |
|---|---|---|---|
| pthread_t | 5 | 400 | 80 |
| pthread_mutex_t | 3 | 267 | 89 |
| pthread_cond_t | 6 | 458 | 76 |
| pthread_barrier_t | 3 | 243 | 81 |
| sem_t | 4 | 329 | 82 |

### 4.3 The Enqueue Operation: Alternative Wakeup Policies

Recall from the introduction (Section 1) that an enqueue operation is needed to implement FIFO and other specialized wakeup policies.

A subtle issue requires the user of McMini to define an additional *enqueue* transition. This is because a "wait" transition (a transition whose visible operation is pthread_mutex_lock, sem_wait, or pthread_cond_wait) consists of two steps. In the first step, the internal state is adjusted to indicate that an additional thread is waiting. In the second step, the thread wakes up and proceeds to execute the remaining invisible operation of the current transition.

Normally, the first step (adjusting the internal state) is independent with all other transitions. Hence, we can consider the two steps as a single, atomic visible operation for the given transition.

An exception occurs when the internal state is made visible to other transitions. An example of this occurs when considering `sem_getvalue`, which returns the internal value (count) of a semaphore. Observe that `sem_getvalue` is independent with a `sem_wait` transition. At the first step, the value is incremented, and at the second step, the value is decremented and the thread is woken up and continues executing.

However, `sem_getvalue` is not independent with first step of `sem_wait`, when viewed in isolation. Hence, we must define an extra transition, `ENQUEUE_sem_wait`. The target program's `sem_wait` is then translated as two transitions:
`ENQUEUE_sem_wait` and `FINISH_sem_wait` The value returned by `sem_getvalue` differs according to whether it is executed before or after a second thread executes `ENQUEUE_sem_wait`. See Listing 3 for an example of how the user of McMini can implement an enqueue operation.

In contrast to FIFO, LIFO and the other policies supported by McMini, other DPOR-based model checkers enforce only the arbitrary wakeup policy. In the case of arbitrary wakeups, the "enqueue" operation is combined with the call to sem_wait as a single atomic transition. In this case, all waiting threads will be enabled after a sem_post operation. In general, FIFO, LIFO and other policies are defined for a specific multi-threaded operation, such as sem_wait. For each multi-threaded operation, any one of its policies can be chosen independently of such choices for any other operation.

◼ **Listing 3** Supporting different wake-up policies for semaphores

```
1  int mc_sem_wait(sem_t *sem) {
2      // Wait until the scheduler tells this thread to enqueue (enter the queue)
3      thread_post_visible_operation_hit<sem_t*>(typeid(MCSemEnqueue), &sem);
4      thread_await_mc_scheduler();
5      // Wait until the scheduler tells this thread to leave the queue
6      thread_post_visible_operation_hit<sem_t*>(typeid(MCSemWait), &sem);
7      thread_await_mc_scheduler();
8      return __native_sem_wait(sem);
9  }
```

The addition of an enqueue operation does not necessarily cause McMini to spend more time in modeling a multithreaded program. This is because the enqueue operation can be modeled as independent of most other operations. For example, sem_wait_enqueue is dependent with sem_post, but it is independent with all other thread operations. Hence, the number of traces (thread schedules) to be explored does not greatly increase. In fact, because FIFO and LIFO offer fewer thread schedules to explore, using a restricted wakeup policy is both more realistic (more representative of the underlying thread library) and also more efficient in time. See Section 5.6 for an experiment showing the effects of different wakeup policies.

## 5 Experimental Evaluation

All experiments were run on: Intel Xeon CPU E5-2690 v3, running at 2.60GHz with Linux distribution CentOS 7.9.2009. All programs were compiled with gcc-8.1.0 and g++-8.1.0 at optimization level -O3.

Each benchmark is executed five times. The mean and sample standard deviation are reported for the time to perform model-checking on each benchmark. Unless otherwise indicated, the ordering of wake ups by McMini was from lowest thread ID to highest. This ensured deterministic results for the number of transitions and traces across different versions of the thread library.

The experiments are designed to answer the following research questions.

**Q1:** Does McMini show performance that is reasonably consistent with other DPOR-based model checkers? (McMini has just 5,000 lines of source code. In order to validate the remaining research questions and answers, one must first show that McMini has "respectable" performance, and is not simply a "toy" system, when compared to existing, highly optimized model checkers.) (See Sections 5.1, and 5.2.)

**Q2:** What is the performance when implementing new thread primitives? (See Sections 5.3 and 5.4.)

**Q3:** One of the novel features of McMini is the ability to implement a new thread operation, that would traditionally be defined as two or more traditional thread primitives. A traditional model checker would need to model this as two or more transitions. How much improvement is obtained by McMini when implementing the new thread operation as a single transition? (See Section 5.7.)

**Q4:** How can this approach support the change in the behavior of the primitives? How well can the McMini approach model unusual behaviors such as spurious wakeup for condition variables, or specific wakeup policies (e.g., FIFO and LIFO). (Traditional DPOR packages do not model these specialized behaviors.) (See Sections 5.5 and 5.6.)

**Q5:** Does the McMini approach work with existing real-world programs? (See Section 5.8.)

### 5.1 Commonly Used Benchmarks for Model Checking

We begin by showing that although McMini is of small size (just 5,000 lines of source code), it has reasonable performance for DPOR-based model checking.

We test McMini with several benchmark programs using standard thread primitives, and we also test the performance of McMini when extended with some custom thread primitives. The tests demonstrate that it can be more efficient to implement a custom thread primitive directly in McMini than to use a library function that defines the custom thread operation in terms of lower-level thread primitives.

Unfortunately, there does not appear to exist any generally recognized benchmarks for model checking, whether as a function of the total number of transitions or the total CPU time expended. The closest thing to such a benchmark appears to be the Dining Philosophers problem. Later, in Figure 2 of Section 5.2, the performance of McMini on the Dining Philosophers problem is shown to be roughly competitive with the state-of-the-art performance from the VeriSoft package [21].

The column names have the following meanings:

**Benchmark:** the name of the program tested.

**Threads (Thrs.):** the number of working threads.

**Transitions (Trans.):** the number of DPOR transitions needed to finish model-checking that particular program.

**Traces:** how many thread schedules or DPOR backtracking points were created by McMini

**Deadlocks (Dlocks.):** the number of traces that lead to a deadlock.

**Time:** the mean time, in seconds, to perform model-checking for that particular benchmark test.

**Standard Deviation (S.D.):** the sample standard deviation for a particular benchmark test.

**Assertion:** the number of the trace at which McMini caught the assertion.

Next, a brief explanation of terminology for the benchmark tests is provided:

- philosophers_*: the classic dining philosophers problem where each thread represents one philosopher. Each philosopher acquires a dining fork (either a semaphore or mutex) and eats once.

- producer_consumer_*: the classic producer-consumer problem where each thread represents one producer or consumer.
- reader_writer_*: the classic reader-writer problem. Each reader/writer acquires the lock exactly once.
- simple_*: test programs whose names start with the prefix "simple" are less complex implementations using the object mentioned in the rest of the program's name.

The '*' character is replaced in each of the benchmark tests by an indication: 1) which synchronization mechanism was used in its implementation; 2) if there is a reader or writer preference in acquiring the object; and 3) whether or not the program can produce a deadlock.

Tables 2, 3, and 4 present the time, number of transitions, and (if applicable) the number deadlocks encountered. They represent programs with no deadlock (Table 2), and programs with a deadlock (Tables 3 and 4). For the latter programs, the time and transitions are shown both to reach the first trace (thread schedule) with a deadlock, and to produce all such traces.

■ **Table 2** Running the programs with no deadlock

| Benchmark | Threads | Trans. | Traces | Time | S.D. |
|---|---|---|---|---|---|
| philosophers_mut | 6 | 37,968 | 2,507 | 11.16 | 0.19 |
| philosophers_sem | 5 | 136,931 | 10,017 | 37.49 | 0.23 |
| reader_writer_fifo | 3 | 935 | 83 | 0.37 | 0.01 |
| reader_writer_mut_reader_preferred | 3 | 9,023 | 831 | 2.76 | 0.04 |
| producer_consumer_mut_sem | 2 | 1156 | 65 | 0.45 | 0.01 |
| simple_barrier_with_threads | 10 | 53 | 1 | 0.04 | 0.00 |

Note that the barrier program in Table 2 is modeled with a single trace. This is because all of the threads in a barrier program are independent.

## 5.2 McMini's Moderation of Combinatorial Explosion

■ **Table 3** Finding the first deadlock (All programs are variants used to create deadlock.)

| Benchmark | Thrs. | Trans. | Traces | Dlocks. | Time | S.D. |
|---|---|---|---|---|---|---|
| simple_cond_broadcast | 10 | 46 | 1 | 1 | 0.04 | 0.00 |
| simple_barrier_with_threads | 5 | 25 | 1 | 1 | 0.02 | 0.00 |
| philosophers_mut | 10 | 37,627 | 2,585 | 1 | 14.38 | 0.06 |
| philosophers_sem | 5 | 7,579 | 513 | 1 | 2.25 | 0.02 |
| producer_consumer_mut_sem | 10 | 53 | 1 | 1 | 0.04 | 0.00 |

Sleep sets were successfully used to moderate the combinatorial explosion. This is illustrated with the Dining Philosophers program, in which each philosopher dines once (see Tables 5 and 6, and its graphical representation in Figure 2).

▪ **Table 4** Finding all deadlocks (All programs are variants used to create deadlock.)

| Benchmark | Threads | Trans. | Traces | Dlocks. | Time | S.D. |
|---|---|---|---|---|---|---|
| simple_cond_broadcast | 5 | 298,010 | 67,680 | 16,680 | 154.70 | 0.57 |
| philosophers_sem | 5 | 157,439 | 11,208 | 120 | 42.97 | 0.27 |
| producer_consumer_mut_sem | 3 | 29,753 | 3,619 | 3,154 | 12.39 | 0.18 |

▪ **Table 5** Dining philosophers comparison with sleep sets (no deadlock)

| Benchmark | Threads | Trans. | Traces | Time | S.D. |
|---|---|---|---|---|---|
| philosophers_mut | 4 | 834 | 45 | 0.26 | 0.00 |
| philosophers_mut | 5 | 5,188 | 311 | 1.38 | 0.01 |
| philosophers_mut | 6 | 37,968 | 2,507 | 11.16 | 0.19 |
| philosophers_mut | 7 | 321,202 | 23,016 | 108.77 | 0.66 |
| philosophers_mut | 8 | 3,097,702 | 236,981 | 1,226.98 | 3.79 |

▪ **Table 6** Dining philosophers comparison without sleep sets (no deadlock)

| Benchmark | Threads | Trans. | Traces | Time | S.D. |
|---|---|---|---|---|---|
| philosophers_mut | 4 | 9,779 | 362 | 2.35 | 0.04 |
| philosophers_mut | 5 | 291,932 | 9,757 | 65.34 | 0.31 |
| philosophers_mut | 6 | 9,772,424 | 303,575 | 19,285.38 | 20.40 |

In Figure 2, we compare McMini's performance on Dining Philosophers with that of VeriSoft [21, Figure 5]. The state-less curve in Figure 5 of VeriSoft [21] refers to classical depth-first search. The curve for Algorithm 2 of VeriSoft [21, Figure 5] refers to an implementation using both sleep sets and persistent sets, though not DPOR. For comparison, McMini implements sleep sets and vector clocks with DPOR. Abdulla et al. [2] note in their introduction that a difference in the order that DPOR explores traces can result in up to an order of magnitude difference in the number of total explored traces. The difference between McMini and VeriSoft can be accounted for by VeriSoft having picked an ordering with fewer sleep-set blocked traces.

## 5.3 Variants of Reader-Writer Locks

The ability of McMini to define new primitives allows one to define multiple variants of reader-writer locks, according to the idiosyncrasies of the target application. An application may implement them in the conventional way, using condition variables. However, McMini can accelerate the speed of model checking by modeling these same multi-threaded constructs as thread primitives in McMini.

Table 7 shows different variants of reader-writer programs, modeled both as condition variables (implemented in the program) and as McMini native primitives (implemented in McMini). Each of the programs has one writer and two readers.

# McMini: A Programmable DPOR-Based Model Checker for Multithreaded Programs

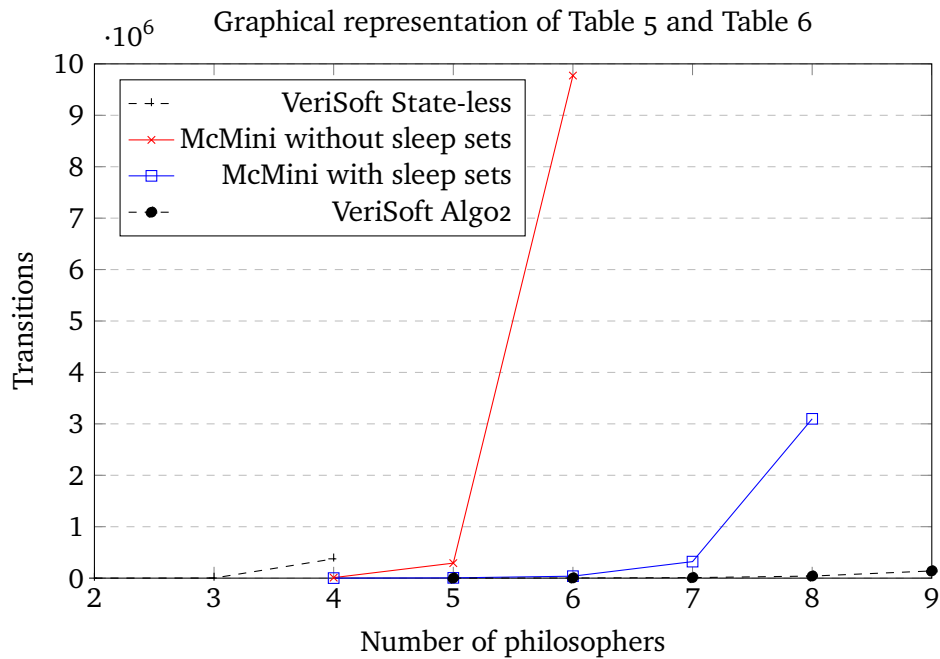

**Figure 2** Scalability of Dining Philosophers (McMini achieves reasonable performance at up to 7 dining philosophers compared to the larger VeriSoft model-checking package used at Lucent Technologies [21, Figure 5].)

**Table 7** 1 Reader, 2 Writers (no deadlock)
(rw_ prefix indicates reader_writer_.)

| Benchmark | Thrs. | Trans. | Traces | Time | S.D. |
|---|---|---|---|---|---|
| rw_cond_no_preference | 3 | 739,912 | 82,540 | 345.99 | 3.30 |
| rw_rwlock_no_preference | 3 | 282,207 | 34,415 | 75.02 | 0.63 |
| rw_cond_reader_preferred | 3 | 604,750 | 53,527 | 251.32 | 1.29 |
| rw_rwlock_reader_preferred | 3 | 284,374 | 36,080 | 83.25 | 0.75 |
| rw_cond_writer_preferred | 3 | 538,118 | 59,884 | 234.94 | 0.50 |
| rw_rwlock_writer_preferred | 3 | 284,374 | 36,080 | 83.40 | 0.54 |

## 5.4 Readers and Two Writers: Modeling a Unique Thread Paradigm as a New Primitive

Next, consider an application that requires only a special case of the general readers-writers problem: multiple readers and just two writers. Each writer thread uses a unique writer lock: pthread_rwwlock_wrlock1 and pthread_rwwlock_wrlock2. The application is assumed to have coded this case, rather than the general case of multiple writers, for performance.

We demonstrate modeling it both with the application's original code, using condition variables, and with a new primitive in McMini. The three threads represent one reader and two writers. Note that the primitive is 1000 times faster.

▪ **Table 8** Readers and Two Writers (no deadlock)

| Benchmark | Threads | Trans. | Traces . | Time | S.D. |
|---|---|---|---|---|---|
| reader_two_writers_cond | 3 | 793,912 | 82,540 | 346.87 | 1.08 |
| reader_writer_rwwlock | 3 | 919 | 109 | 0.34 | 0.01 |

## 5.5 Support for Spurious Wakeups

We demonstrate support for spurious wakeups. Consider Listing 4, below. The listing shows a standard implementation of producer-consumer using condition variables.

A classic bug in the use of condition variables occurs when the '**while**' in the consumer() function of Listing 4 is replaced by '**if**'. In this case, a spurious wakeup produces a bug that allows the consumer to attempt consumption before any producers have produced anything.

▪ **Listing 4** Erroneous producer-consumer using condition variables (The use of '**if**' in place of '**while**' in the code creates a classic bug that is exhibited *only* when spurious wakeups occur.)

```
1  int number_consumed = 0;
2  int num_elements_to_consume = 0;
3  void do_consume() {
4    mutex_lock
5    if (num_elements_to_consume == 0) {
6      pthread_cond_wait(...);
7    }
8    mutex_unlock
9    ... consume it ...
10   mutex_lock
11   assert(num_elements_to_consume > 0);
12   num_elements_to_consume--;
13   mutex_unlock
14 }
15 void consumer() {
16   while (1) {
17     do_consume();
18     number_consumed++;
19   }
20 }
21 void producer() {
22   while (1) {
23     mutex_lock
24     num_elements_to_consume++;
25     pthread_cond_signal(...);
26     mutex_unlock
27   }
28 }
```

Table 9 shows the results of running McMini on Listing 4 when using '**while**' (no bug) and '**if**' (bug in presence of spurious wakeups) with a single consumer and increasingly more producers. Each thread was restricted to run at most six transitions in every thread schedule to account for the infinite loops. For the sake of efficiency, spurious wakeups are modeled to allow at most one spurious wakeup on each condition variable.

Note that the depth-first nature of DPOR implies that the first thread chosen will execute transitions until blocked. Hence, if the consumer thread is chosen first, then McMini detects an assertion failure on the very first trace. In order to show a less trivial result, the programs of Table 9 were implemented to spawn all producer threads before spawning the consumer thread. The –max-thread-depth flag was set to limit the number of transitions by a producer, so that the consumer thread would eventually be spawned.

■ **Table 9** Producer-Consumer Spurious Wakeup (The number of threads in parentheses indicates the number of producer threads '+' the number of consumer threads.)

| Benchmark | Threads | Trans. | Traces | Assertion | Time | S.D. |
|---|---|---|---|---|---|---|
| no_spurious | 2 (1+1) | 84 | 21 | - | 0.09 | 0.04 |
| spurious | 2 (1+1) | - | - | @ TRACE 28 | 0.04 | 0.00 |
| no_spurious | 3 (2+1) | 2,301 | 907 | - | 1.41 | 0.03 |
| spurious | 3 (2+1) | - | - | @ TRACE 1,046 | 0.47 | 0.04 |
| no_spurious | 4 (3+1) | 101,509 | 48,636 | - | 83.42 | 0.22 |
| spurious | 4 (3+1) | - | - | @ TRACE 45,040 | 17.58 | 0.13 |

**5.6 Different Wakeup Policies and Enqueue**

Table 10 shows the number of traces and transitions exhibited by McMini while modeling the dining philosophers program using semaphores with three threads. Five different wakeup policies for semaphores shown here: FIFO, LIFO, and arbitrary with three different choices for an enqueue operation. Note that LIFO has fewer transitions than FIFO. This is because there are fewer enabled transitions interacting with the semaphore. Thus McMini backtracks less often.

A user of McMini who implements a particular thread wakeup policy would need to consider, for correctness, with what other thread operations an enqueue operation should be dependent. Note that in the case of the "arbitrary" wakeup policy, it is possible to integrate the implementation of the enqueue operation into the sem_wait operation.

1. For a semaphore waking threads using a FIFO wakeup policy, the enqueue operation is dependent with another enqueue on the same semaphore
2. For a semaphore waking threads using a LIFO wakeup policy, the enqueue operation is dependent with with another enqueue on the same semaphore and a `sem_wait` on the same semaphore
3. For a semaphore waking threads using an arbitrary wakeup policy, the enqueue operation is independent with all other operations
4. For a semaphore waking threads using an arbitrary wakeup policy, the enqueue operation is dependent with another enqueue on the same semaphore

5. For a semaphore waking threads using an arbitrary wakeup policy without an enqueue operation. The enqueue is atomic within the sem_wait

■ **Table 10** Difference wake-up policies for semaphores

| Benchmark | Trans. | Traces |
| --- | --- | --- |
| LIFO | 81,282 | 4,519 |
| FIFO | 104,216 | 5,801 |
| Arbitrary (with independent enqueue) | 69,924 | 3,918 |
| Arbitrary (with dependent enqueue) | 11,344 | 203,003 |
| Arbitrary (without enqueue) | 10,840 | 601 |

The more restrictive wakeup policies (FIFO and LIFO) have few traces to explore, since for a given semaphore, at most one thread calling sem_wait on that semaphore will be enabled. Nevertheless, the arbitrary case makes up for this if the enqueue operation is independent with *all* other operations. In that case, the DPOR algorithm reduces the search space. Finally, if the enqueue operation is dependent with some other thread operations, as described above, then the number of traces grows very large.

As discussed elsewhere, restricted wakeup policies such as FIFO and LIFO are sometimes important for more accurately modeling the wakeup policy of the underlying thread library.

## 5.7 Library-Defined versus McMini-Defined Thread Operations

An advantage of McMini is the ability to directly model multithreaded operations as primitives, instead of modeling an implementation of the primitive (e.g., using condition variables). This becomes a potential inefficiency in model checking, when an end user must implement a more complex multi-threaded constructs through multiple thread operations.

In order to capture the cost of this inefficiency, we take three cases where we assume that the desired thread primitive does not exist in the McMini user's thread library. We implement each of the thread operations through an end-user library that implements the desired thread operation inside the target application. And we implement the thread operation directly as a new McMini primitive.

A small library is created to compare McMini's performance of programs with semaphores and condition variable. Table 11 shows a comparison of the number of transitions McMini ran when using condition variables based directly on the native pthread library or on an implementation based on semaphores.

Intuitively, the semaphore-based implementation turns what would be an atomic operation carried out by a single visible operation into several visible operations. The implementation of pthread_cond_wait(), for example, consists of five visible operations on mutexes and semaphores, as opposed to two visible operations when modeled directly as (native) condition variables. McMini can be extended to support such atomic operations to reduce the number of visible operations.

Next, a brief explanation of the programs in Table 11 follows.

- simple_barrier: The library implementation uses a custom barrier defined in our small library which uses a mutex, semaphore, and a counter to implement it. When a thread enters the barrier it waits on the semaphore. The last thread that enters the barrier posts to everyone else. The McMini primitive uses a barrier and was run with an arbitrary wake-up policy.
- simple_cond_broadcast: The library implementation uses a custom condition variable defined in our small library which uses a mutex and a semaphore to implement it. The McMini primitive uses a condition variable broadcast and was run using FIFO wake-up policy.
- philosophers_semaphore: The library implementation uses a custom semaphore defined in our small library which uses a mutex and a condition variable to implement it. The McMini primitive uses the semaphore as a "fork" in the dining philosophers problem, and was run using the FIFO wake-up policy.

■ **Table 11** Comparing the number of transitions for McMini to model-check simple programs using McMini-defined primitives versus library-based implementations using condition variables; all three programs use two threads.

| Benchmark | Library Implementation | McMini Primitive |
| --- | --- | --- |
| simple_barrier | 114 | 90 |
| simple_cond_broadcast | 4029 | 28 |
| philosophers_semaphore | 2873 | 176 |

## 5.8 Detecting a Data Race in a Real-World Program

As a demonstration of McMini's usability on software unrelated to the current authors, McMini has been used to identify a data race in a real-world program, pthread_pool [20], by Jon Gjengset at MIT. The target program provides thread pools on top of POSIX threads. The shared variables (fields) of the struct pool referenced by pool_start, pool_end and pool_enqueue, pool_wait, thread were all annotated with a McMini macro MC_READ/MC_WRITE. The data race occurs when a thread terminates the pool (see pool_end() in the source), with running worker threads without first waiting on the worker threads (see pool_wait()). The data race could cause worker threads to erroneously continue work even after cancellation, even if no such work remains. McMini is thus capable of detecting data races in more complicated programs.

## 6 Related Work

There are many approaches to creating a Model Checking software package [12]. The current work concentrates on the use of DPOR (Dynamic Partial Order Reduction), first introduced in the fundamental work of Flanagan and Godefroid [17]. See Godefroid and Yen [22, Section 2.5] for an excellent overview of this DPOR approach.

Yang et al. [42] showed an early implementation of DPOR-based model checking that supports mutex, semaphore, condition variable. Unlike the current work, their implementation is not customizable: they demonstrate their work for a single policy on wakeups only and do not document which policy is being used. For example, if their policy includes a possible FIFO ordering of wakeups for producer-consumer, then this could lead to starvation for a particular thread, but this is not discussed. Still earlier, Yang et al. had demonstrated a parallelized version of DPOR-based model checking [41].

Similarly, Zaks et al. [44] demonstrated using the SPIN model checker [24] to verify multi-threaded C programs that support the mutexes, semaphores and condition variables of the pthreads library. That work uses the Promela model of SPIN and some embedded portions of C code, in order to check an LTL logic property. By using LTL, they can employ algorithmic guarantees for checking properties such as progress or liveness (absence of starvation), unlike the heuristic approach of the current work. They use a tool "pancam" to implement a virtual machine that executes LLVM byte code. Like the current work, they use *context-bounded model checking* [32, 35] to limit the number of context switches. (However, the current work places a limit on the number of context switches *per thread*.)

In addition to SPIN, DIVINE [7, 8] and SimGrid [9] also support verification of liveness properties through temporal logics (e.g., LTL), encoded in Promela and using an underlying theory of Büchi Automata [19, 38]. Kripke structures are sometimes also used, to optimize for symmetries [16, 39]. In distinction from temporal logics, the approach of this work to starvation detection is heuristic. McMini is able to provide the same theoretical guarantees as LTL for detecting starvation only if its threadMaxLimit and extraLivenessTransitions parameters are set to be sufficiently large.

Zaks et al. employ SPIN on C programs [44]. They take advantage of the compiled LLVM byte code of the target application and SPIN, to consider all interleavings of reads and writes among shared variables. While multithreaded operations are seen at assembly level, they are treated as an atomic operation, called a *superstep*. Their supersteps are roughly comparable to the *transitions* used here. The work also employs DPOR [17]. Their experiments with the dining philosophers problem is limited to 5 philosophers, involving 25,021 supersteps.

Another notable work is the CHESS project [23] of Microsoft. This includes a DPOR-based model checker [33] to reproduce "Heisenbugs" (rare race conditions). They also directly model multi-threaded programs, rather than explicitly supporting the common POSIX synchronization operations for mutex, semaphore and condition variable. They also limit the number of thread context switches in a trace, but they do so using a best-first search heuristic to emphasize regions of code with likely race conditions. That work is extended in GAMBIT [13, Section 4.5], which employs different priority functions for preemption-bounded search.

Finally, Sthread [14] emphasizes an *in-vivo model checker*, employing the S4U interface [37] of SimGrid [9]. It is in vivo in the sense that it directly executes (many times) on the compiled binary of the target code, while using a scheduling semaphore to force different thread schedules. It is "modeling" the target application in the sense that it interposes on multi-threaded operations and "models" the changing state of a thread (e.g., wakeup, sleeping while waiting for for a particular event). However, the target code outside of the thread operations is executed directly (in vivo), and is not modeled. Two other examples of in-vivo model checking are the current work (McMini) and Yang et al. [40]. In addition, Zaks et al. [44] uses an in-vivo strategy on top of SPIN [24], in which they directly execute an intermediate version of the compiled target code.

There have also been numerous variants of DPOR, including those which are context-sensitive [3], data-centric [10], optimal according to a theoretical criterion [1, 5], optimal in the sense of maximal causality reduction [25], and stateful [43] (in order to avoid re-exploring visited states). In [36], the authors modify the DPOR algorithm to take advantage of commutativity of certain visible operations, such as reading a shared variable by two different threads. Other lines of work explore DPOR in the context of constraints [4] and conditional explorations [28].

An older work, VeriSoft [21], operates directly on C programs, but uses static partial order reduction.

## 7  Conclusion and Future Work

An extensible, in vivo model checker was demonstrated. The in vivo feature allows the user of McMini [30] to model additional thread operations, and also to customize existing thread operations, by directly executing the target program's code over many schedules (traces). Defining new primitive thread operations for McMini allows one to directly model the multithreaded paradigms found in applications. This is more efficient than the traditional approach of modeling the underlying primitives (e.g., condition variables) that were used to implement the application-specific paradigm. This approach was shown to be up to five times faster than traditional implementations on top of condition variables.

The ability to model-check variants of standard wakeup policies has also been demonstrated. This requires an additional *enqueue* operation before each "wait" transition (e.g., sem_wait, pthread_cond_wait). The result enables McMini to model the actual wakeup policies of the underlying thread library: Whether the sem_post or signal arrived before or after the thread had entered the "wait" transition. This matters in, for example, distinguishing FIFO from LIFO. This specificity of behavior also often allows McMini to more efficiently execute model checking, than for for the default case of arbitrary wakeups by any thread.

Finally, the ability to model spurious wakeups for condition variables represents an important advance, not found in other DPOR-based model checkers. In a simple producer-consumer program, the ability to model spurious wakeups allowed McMini to discover a bug that would not be found in traditional DPOR-based model checkers.

**Acknowledgements**   The authors are grateful for the careful reading and comments by the reviewers, which greatly improved the exposition.

## A  Defining a New McMini Thread Operation

This appendix provides a running example of defining a new thread operation in McMini. It demonstrates how to add support for the pthread_rwlock_rdlock() function in McMini for writer-preferred, reader-writer locks with a FIFO wake-up ordering to highlight key points in the process of adding new transitions.

While this implementation assumes writer-preferred and a FIFO queue of waiting threads, it will be clear how to alter the description to McMini to, for example, employ reader-preferred or an arbitrarily ordered queue of waiting threads. The relevant changes to this implementation would primarily be in Listing 11, below.

### A.1  Defining Reader-Writer Locks as McMini Primitives

Listing 5 defines a McMini wrapper function that is invoked when the target application calls pthread_rwlock_rdlock(). Listing 6 then defines the object our new transitions operate on, while Listings 8, 9, and 10 show the implementations of the three essential attributes for any transition. The three attributes follow: *enabled* (Listing 8); *dependent* (Listing 9); and *coenabled* (Listing 10). Finally, Listing 11 shows the *action* to be taken in the thread trace when the current operation is executed.

The three attributes are implemented to provide "hints" to the DPOR algorithm about how to reduce the number of thread schedules that need to be searched.

The action involves modifying McMini's internal state inside the scheduler to reflect the fact that a particular transition was executed. Typically, for each action McMini executes, it may emulate the semantics entirely in McMini, or in some cases it can simply invoke the corresponding underlying native thread library (e.g., libpthread.so) to execute the transition represents.

#### A.1.1  Define a Wrapper Function for pthread_rwlock_rdlock()

We begin by defining a wrapper function for each thread operation to be modeled in the target application. Recall the wrapper function employs McMini-internal semaphores, to enable a protocol similar to producer-consumer, with the user thread invoking a thread operation as a producer, and the scheduler from Figure 1 acting as the consumer. This allows the scheduler to be informed of each transition in the current schedule trace.

A McMini wrapper for pthread_rwlock_rdlock() is shown below.

■ **Listing 5**  A wrapper function for pthread_rwlock_rdlock()

```
int mc_pthread_rwlock_rdlock(pthread_rwlock_t *rwlock) {
  MCRWLockShadow shadowLock(rwlock);
  …
  thread_post_visible_operation_hit<MCRWLockShadow>(
    typeid(MCRWLockReaderLock), &shadowLock);
  thread_await_mc_scheduler();
  return __native_pthread_rwlock_rdlock(rwlock);
}
```

# McMini: A Programmable DPOR-Based Model Checker for Multithreaded Programs

### A.1.2 Define a Visible Object for RWLock

Listing 6 then defines MCRWLock as a subtype of MCVisibleObject. The MCVisibleObject type keeps track of any state needed to correctly emulate the primitive's unique behavior. Three queues are added to keep track of active readers and waiting readers and writers. The terminology "visible object" is used by analogy with "visible operations" (see Section 2).

**Listing 6** Define a RWLock object (for read/write operations)

```cpp
#include "mcmini/MCVisibleObject.h"
struct MCRWLock : public MCVisibleObject {
  MCOptional<tid_t> active_writer = MCOptional<tid_t>::nil();
  std::vector<tid_t> active_readers;
  std::queue<tid_t> reader_queue;
  std::queue<tid_t> writer_queue;
  …
};
```

### A.1.3 Specify the Three Attributes Required by DPOR

Each transition added to McMini inherits from MCTransition and modifies the state of whatever primitive it acts on appropriately (i.e. according to the semantics of the operation). The transition's dependency conditions and co-enabled relations with any existing visible operations should be specified, as well as any conditions affecting the whether the transition is enabled. Recall that by default, McMini treats a transition as dependent/co-enabled with all other transitions.

**Listing 7** The Reader Operation of a RWLock: Three Attributes and an Action

```cpp
#include "mcmini/MCTransition.h"
struct MCRWLockReaderLock : public MCTransition {
  MCRWLock *rwlock;
  // Attribute 1: Must be overridden
  bool enabledInState(const MCState *) const override;
  // Attribute 2: Default returns true always, i.e. is dependent
  bool dependentWith(const MCTransition *) const override;
  // Attribute 3: Default returns true always, i.e. everything is co-enabled
  bool coenabledWith(const MCTransition *) const override;
  // Action: Default does nothing
  bool applyToState(MCState *) override;
};
```

**Listing 8** Reader Operation: Attribute 1: When is a thread enabled?

```cpp
bool MCRWLockReaderLock::enabledInState(const MCState *state) const {
  return this->rwlock->canAcquireAsReader(this->getThreadId());
}
…
bool MCRWLock::canAcquireAsReader(tid_t tid) const {
  // Assuming writer preferred and the reader queue is FIFO
  return !hasEnqueuedWriters() && !isWriterLocked() && reader_queue.front() == tid;
}
```

To support pthread_rwlock_rdlock(), two transitions are defined: MCRWLockReader-Lock and MCRWLockReaderLockEnqueue. The latter transition (enqueue) is needed only to define a particular wakeup policy. As discussed in section 4.2, only a single transition subclass is needed for each wrapper. Listings 8, 9, and 10 show the implementations of these three essential methods for MCRWLockReaderLock.

■ **Listing 9** Reader Operation: Attribute 2: What other thread operations are dependent (or independent) with this operation?

```cpp
bool MCRWLockReaderLock::dependentWith(const MCTransition *other) const {
  const MCRWLockReaderLock *maybeRWLockReaderOperation =
    dynamic_cast<const MCRWLockReaderLock *>(other);
  if (maybeRWLockReaderOperation) {
    return false;
  }
  const MCRWLockTransition *maybeRWLockOperation =
    dynamic_cast<const MCRWLockTransition *>(other);
  if (maybeRWLockOperation) {
    return *maybeRWLockOperation->rwlock == *this->rwlock;
  }
  return false;
}
```

■ **Listing 10** Reader Operation: Attribute 3: What thread operations are coenabled with this?

```cpp
bool MCRWLockReaderLock::coenabledWith(const MCTransition *other) const {
  const MCRWLockReaderLock *maybeReaderLock =
    dynamic_cast<const MCRWLockReaderLock *>(other);
  if (maybeReaderLock) {
    return true;
  }
  const MCRWLockWriterLock *maybeWriterLock =
    dynamic_cast<const MCRWLockWriterLock *>(other);
  if (maybeWriterLock) {
    return *maybeWriterLock->rwlock != *this->rwlock;
  }
  return true;
}
```

**McMini: A Programmable DPOR-Based Model Checker for Multithreaded Programs**

### A.1.4 Specify the Action of the Reader Operation for pthread_rwlock_rdlock()

Finally, we modify McMini's state according to the behavior of pthread_rwlock_rdlock() in Listing 11:

■ **Listing 11** Reader Operation: Action: Update RWLock for pthread_rwlock_rdlock

```cpp
// Default: Does nothing
bool MCRWLockReaderLock::applyToState(MCState *state) {
  this->rwlock->reader_lock(this->getThreadId());
}
…
void MCRWLock::reader_lock(tid_t tid) {
  // Assuming writer preferred and the reader queue is FIFO
  this->active_readers.push_back(tid);
  this->reader_queue.pop();
}
```

### A.2 Adding the Enqueue Operation for pthread_rwlock_rdlock()

To complete support for pthread_rwlock_rdlock(), we add the MCRWLockReaderLock-Enqueue transition that enqueues a thread as a reader. See Section 4.3 for a more detailed description of the implementation of enqueue operations and why they are useful.

The pthread_rwlock_rdlock() function employs an enqueue operation, followed by acquiring the read lock. If the two operations were combined into a single operation, as shown in the previous sections, we wouldn't correctly mark that a reader thread were waiting to acquire the lock. This would affect the behavior of a reader-preferred reader-writer lock, where readers should be scheduled before writers.

The Enqueue operation is analogous to MCRWLockReaderLock:

■ **Listing 12** Enqueue for RWLock: Three attributes

```cpp
#include "mcmini/MCTransition.h"
struct MCRWLockReaderEnqueue : public MCRWLockTransition {
  MCRWLock *rwlock;
  // Attribute 1: Default: Enabled by default, must override
  bool enabledInState(const MCState *) const override;
  // Attribute 2: Default: Everything is dependent
  bool dependentWith(const MCTransition *) const override;
  // Attribute 3: Default: Does nothing
  bool applyToState(MCState *) override;
};
```

We define dependency conditions as before. Note that an enqueue operation is always co-enabled with other transitions and is itself always enabled, so the default suffices here:

■ **Listing 13** Enqueue for RWLock: Specify dependent operations

```
bool MCRWLockReaderEnqueue::dependentWith(const MCTransition *other) const {
  const MCRWLockTransition *maybeRWLockOperation =
    dynamic_cast<const MCRWLockTransition *>(other);
  if (maybeRWLockOperation) {
    return *maybeRWLockOperation->rwlock == *this->rwlock;
  }
  return false;
}
```

Finally, we enqueue the reader into the lock's reader queue when applied:

■ **Listing 14** Enqueue: Specify the action

```
void MCRWLockReaderEnqueue::applyToState(MCState *state) {
  // Enqueue this thread as a reader
  this->rwlock->enqueue_as_reader(this->getThreadId());
}
…
void MCRWLock::enqueue_as_reader(tid_t tid) {
  this->reader_queue.push(tid);
}
```

### A.3 Extending the Example: Many Readers and Two Writers

Suppose we wanted to implement a new primitive pthread_rww_lock_t that behaved the same as the classic pthread_rw_lock_t except that there were two *types* of writers: writer 1 and writer 2. A thread can acquire the lock as either type of writer by making calls to pthread_rwwlock_wrlock1() and pthread_rwwlock_wrlock2() respectively.

Such a primitive can be added with minimal changes to McMini's existing implementation for pthread_rw_lock_t. For simplicity, we assume that we favor either type of writer over readers and have no preference between writers of different types. The following modifications can be made to support the new primitive:

1. One splits the writers_queue into writers1_queue and writers2_queue. The writer threads acquire locks via pthread_rwwlock_wrlock1() and pthread_rwwlock_wrlock2(), respectively.
2. One adds two new transitions: MCRWLockWriter2Enqueue and MCRWLockWriter2Lock. Here, one can copy the MCRWLockWriter2Enqueue transitions for both the first and second writers.

The pseudocode from Listing 8 remains the same, and the check for "are there writers enqueued?" now simply checks both writers1_queue and writers2_queue. Dependency information is also largely unchanged.

Note that such a hypothetical lock, pthread_rww_lock_t, does not need to be implemented natively in the program source or in a library (using e.g. semaphores and mutexes) if the goal is only to perform model checking under McMini. Hence, if an application defines a new type of thread operation, then McMini allows the developer the option of directly testing the new thread operation by encoding its semantics into McMini, instead of modeling the lower-level implementation of the primitive. One can thus test a proposed thread primitive even before implementing it as a function inside the application. This works because McMini allows the developer to write a McMini wrapper function that will be called when testing with McMini.

## References

[1] Parosh Abdulla, Stavros Aronis, Bengt Jonsson, and Konstantinos Sagonas. Optimal dynamic partial order reduction. *ACM SIGPLAN Notices*, 49(1):373–384, 2014. doi:10.1145/2578855.2535845.

[2] Parosh Aziz Abdulla, Stavros Aronis, Bengt Jonsson, and Konstantinos Sagonas. Source sets: a foundation for optimal dynamic partial order reduction. *Journal of the ACM (JACM)*, 64(4):1–49, 2017. doi:10.1145/3073408.

[3] Elvira Albert, Puri Arenas, María García de la Banda, Miguel Gómez-Zamalloa, and Peter J Stuckey. Context-sensitive dynamic partial order reduction. In *International Conference on Computer Aided Verification*, pages 526–543. Springer, 2017. doi:10.1007/978-3-319-63387-9_26.

[4] Elvira Albert, Miguel Gómez-Zamalloa, Miguel Isabel, and Albert Rubio. Constrained dynamic partial order reduction. In *International Conference on Computer Aided Verification*, pages 392–410. Springer, 2018. doi:10.1007/978-3-319-96142-2_24.

[5] Stavros Aronis, Bengt Jonsson, Magnus Lång, and Konstantinos Sagonas. Optimal dynamic partial order reduction with observers. In *International Conference on Tools and Algorithms for the Construction and Analysis of Systems*, pages 229–248. Springer, 2018. doi:10.1007/978-3-319-89963-3_14.

[6] Remzi H. Arpaci-Dusseau and Andrea C. Arpaci-Dusseau. *Operating Systems: Three Easy Pieces*. https://pages.cs.wisc.edu/~remzi/OSTEP/, 2022. ISBN: 978-1985086593.

[7] Zuzana Baranová, Jiří Barnat, Katarína Kejstová, Tadeáš Kučera, Henrich Lauko, Jan Mrázek, Petr Ročkai, and Vladimír Štill. Model checking of C and C++ with DIVINE 4. In *International Symposium on Automated Technology for Verification and Analysis*, volume 10482 of *LNCS*, pages 201–207. Springer, 2017. doi:10.1007/978-3-319-68167-2_14.

[8] Jiri Barnat, Lubos Brim, and Petr Ročkai. DiVinE multi-core–a parallel LTL model-checker. In *International Symposium on Automated Technology for Verification and Analysis*, pages 234–239. Springer, 2008. doi:10.1007/978-3-540-88387-6_20.

[9] Henri Casanova, Arnaud Giersch, Arnaud Legrand, Martin Quinson, and Frédéric Suter. Versatile, scalable, and accurate simulation of distributed applications and platforms. *Journal of Parallel and Distributed Computing*, 74(10):2899–2917, June 2014. doi:10.1016/j.jpdc.2014.06.008.

[10] Marek Chalupa, Krishnendu Chatterjee, Andreas Pavlogiannis, Nishant Sinha, and Kapil Vaidya. Data-centric dynamic partial order reduction. *Proceedings of the ACM on Programming Languages*, 2(POPL):1–30, 2017. doi:10.1145/3158119.

[11] Edmund M. Clarke, E. Allen Emerson, and Joseph Sifakis. Model checking: algorithmic verification and debugging. *Communications of the ACM*, 52(11):74–84, 2009. doi:10.1145/1592761.

[12] Edmund M. Clarke, Thomas A. Henzinger, and Helmut Veith. Introduction to model checking. In *Handbook of Model Checking*, pages 1–26. Springer, 2018. doi:10.1007/978-3-319-10575-8_1.

[13] Katherine E. Coons, Sebastian Burckhardt, and Madanlal Musuvathi. GAMBIT: Effective unit testing for concurrency libraries. *ACM Sigplan Notices*, 45(5):15–24, 2010. doi:10.1145/1837853.1693458.

[14] Gene Cooperman and Martin Quinson. Sthread: In-vivo model checking of multithreaded programs. *The Art, Science, and Engineering of Programming*, 4(3):13:1–13:26, February 2017. doi:10.22152/programming-journal.org/2020/4/13/.

[15] Lucas Cordeiro. SMT-based bounded model checking for multi-threaded software in embedded systems. In *2010 ACM/IEEE 32nd International Conference on Software Engineering*, volume 2, pages 373–376. IEEE, 2010. doi:10.1145/1810295.1810396.

[16] Alastair Donaldson, Alice Miller, and Muffy Calder. Spin-to-Grape: A tool for analysing symmetry in Promela models. *Electronic Notes in Theoretical Computer Science*, 139(1):3–23, 2005. doi:10.1016/j.entcs.2005.09.007.

[17] Cormac Flanagan and Patrice Godefroid. Dynamic partial-order reduction for model checking software. *ACM Sigplan Notices*, 40(1):110–121, 2005. doi:10.1145/1040305.1040315.

[18] Mikhail R. Gadelha, Felipe R. Monteiro, Jeremy Morse, Lucas C. Cordeiro, Bernd Fischer, and Denis A. Nicole. ESBMC 5.0: an industrial-strength C model checker. In *Proceedings of the 33rd ACM/IEEE International Conference on Automated Software Engineering*, pages 888–891, 2018. doi:10.1145/3238147.3240481.

[19] Paul Gastin and Denis Oddoux. Fast LTL to Büchi automata translation. In *International Conference on Computer Aided Verification*, pages 53–65. Springer, 2001. doi:10.1007/3-540-44585-4_6.

[20] Jon Gjengset. A small implementation of thread pools using POSIX threads. https://github.com/jonhoo/pthread_pool, 2013. last accessed May, 2023.

[21] Patrice Godefroid. Software model checking: The VeriSoft approach. *Formal Methods in System Design*, 26(2):77–101, 2005. doi:10.1007/s10703-005-1489-x.

[22] Patrice Godefroid and Koushik Sen. Combining model checking and testing. In *Handbook of Model Checking*, pages 613–649. Springer, 2018. doi:10.1007/978-3-319-10575-8_19.

[23] Peli de Halleux, Madan Musuvathi, Sebastian Burckhardt, and Thomas Ball. CHESS: Find and reproduce Heisenbugs in concurrent programs. https://www.microsoft.com/en-us/research/project/chess-find-and-reproduce-heisenbugs-in-concurrent-programs/, 2008. last accessed May, 2023.

[24] Gerard J. Holzmann. The model checker SPIN. *IEEE Transactions on Software Engineering*, 23(5):279–295, 1997. doi:10.1109/32.588521.

[25] Jeff Huang. Stateless model checking concurrent programs with maximal causality reduction. *ACM SIGPLAN Notices*, 50(6):165–174, 2015. doi:10.1145/2737924.2737975.

[26] Omar Inverso, Truc L. Nguyen, Bernd Fischer, Salvatore La Torre, and Gennaro Parlato. Lazy-CSeq: A context-bounded model checking tool for multi-threaded C-programs. In *2015 30th IEEE/ACM International Conference on Automated Software Engineering (ASE)*, pages 807–812. IEEE, 2015. doi:10.1109/ASE.2015.108.

[27] Omar Inverso and Catia Trubiani. Parallel and distributed bounded model checking of multi-threaded programs. In *Proceedings of the 25th ACM SIGPLAN Symposium on Principles and Practice of Parallel Programming*, pages 202–216, 2020. doi:10.1145/3332466.3374529.

[28] Miguel Isabel. Conditional dynamic partial order reduction and optimality results. In *Proceedings of the 28th ACM SIGSOFT International Symposium on Software Testing and Analysis*, pages 433–437, 2019. doi:10.1145/3293882.3338987.

[29] Anil Kumar Karna, Yuting Chen, Haibo Yu, Hao Zhong, and Jianjun Zhao. The role of model checking in software engineering. *Frontiers of Computer Science*, 12(4):642–668, 2018. doi:10.1007/s11704-016-6192-0.

[30] McMini Team. McMini: A small, extensible DPOR-based model checker. https://github.com/mcminickpt/mcmini, 2023. last accessed May, 2023.

[31] Jeremy Morse, Lucas Cordeiro, Denis Nicole, and Bernd Fischer. Context-bounded model checking of LTL properties for ANSI-C software. In *International Conference on Software Engineering and Formal Methods*, pages 302–317. Springer, 2011. doi:10.1007/978-3-642-24690-6_21.

[32] Madanlal Musuvathi and Shaz Qadeer. Iterative context bounding for systematic testing of multithreaded programs. In *ACM Sigplan Notices*, volume 42, pages 446–455. ACM, 2007. doi:10.1145/1273442.1250785.

[33] Madanlal Musuvathi, Shaz Qadeer, Thomas Ball, Gerard Basler, Piramanayagam Arumuga Nainar, and Iulian Neamtiu. Finding and reproducing Heisenbugs in concurrent programs. In *USENIX Symposium on Operating Systems Design and Implementation (OSDI'08)*, volume 8, 2008.

[34] Chris Newcombe, Tim Rath, Fan Zhang, Bogdan Munteanu, Marc Brooker, and Michael Deardeuff. How Amazon web services uses formal methods. *Communications of the ACM*, 58(4):66–73, 2015. doi:10.1145/2699417.

[35] Shaz Qadeer and Jakob Rehof. Context-bounded model checking of concurrent software. In *International Conference on Tools and Algorithms for the Construction and Analysis of Systems*, pages 93–107. Springer, 2005. doi:10.1007/978-3-540-31980-1_7.

[36] Olli Saarikivi, Kari Kähkönen, and Keijo Heljanko. Improving dynamic partial order reductions for concolic testing. In *2012 12th International Conference on Application of Concurrency to System Design*, pages 132–141. IEEE, 2012. doi:10.1109/ACSD.2012.18.

[37] SIMGRID Team. The S4U interface of SimGrid. https://simgrid.org/doc/latest/app_s4u.html, 2022. last accessed May, 2023.

[38] Fabio Somenzi and Roderick Bloem. Efficient büchi automata from ltl formulae. In *International Conference on Computer Aided Verification*, pages 248–263. Springer, 2000. doi:10.1007/10722167_21.

[39] Wikipedia. Kripke structure (model checking). https://en.wikipedia.org/wiki/Kripke_structure_(model_checking), 2022. last accessed May, 2023.

[40] Junfeng Yang, Tisheng Chen, Ming Wu, Zhilei Xu, Xuezheng Liu, Haoxiang Lin, Mao Yang, Fan Long, Lintao Zhang, and Lidong Zhou. MODIST: Transparent model checking of unmodified distributed systems. In *NSDI'09*, pages 213–228, 2009.

[41] Yu Yang, Xiaofang Chen, Ganesh Gopalakrishnan, and Robert M. Kirby. Distributed dynamic partial order reduction based verification of threaded software. In *International SPIN Workshop on Model Checking of Software*, pages 58–75. Springer, 2007. doi:10.1007/978-3-540-73370-6_6.

[42] Yu Yang, Xiaofang Chen, Ganesh Gopalakrishnan, and Robert M. Kirby. Runtime model checking of multithreaded C/C++ programs. Computer Science technical report UUCS-07-008.pdf, U. of Utah, 2007.

[43] Yu Yang, Xiaofang Chen, Ganesh Gopalakrishnan, and Robert M. Kirby. Efficient stateful dynamic partial order reduction. In *International SPIN Workshop on Model Checking of Software*, pages 288–305. Springer, 2008. doi:10.1007/978-3-540-85114-1_20.

[44] Anna Zaks and Rajeev Joshi. Verifying multi-threaded C programs with SPIN. In *International SPIN Workshop on Model Checking of Software*, pages 325–342. Springer, 2008. doi:10.1007/978-3-540-85114-1_22.

## About the authors

**Maxwell Pirtle** is a fourth-year undergraduate computer science student at Northeastern University in Boston, USA. He is interested in computer systems and formal verification. He is currently a research intern at Inria in Rennes, France, working on the SimGrid project. Contact him at pirtle.m@northeastern.edu.
https://orcid.org/0009-0000-0466-3758

**Luka Jovanovic** is a first-year M.S. student in computer science at the HPC laboratory in Northeastern University in Boston, USA. He is interested in computer systems, machine learning, and theory. Contact him at jovanovic.l@northeastern.edu.
https://orcid.org/0009-0000-1692-1523

**Gene Cooperman** is a professor of Computer Science at Northeastern University in Boston, USA. He had joined GTE Laboratories in 1980, had joined Northeastern University in 1986, and has been a full professor since 1992. He has over 125 refereed publications. He has interests in computer systems, high performance computing, transparent checkpoint-restart, and novel methods for debugging. Contact him at gene@ccs.neu.edu.
https://orcid.org/0000-0003-2175-3848